\documentclass[12pt,prd,ams,twocolumn,eprint,floats,showpacs,tightenlines]{revtex4}

\usepackage{amsmath}
\usepackage{amsfonts}
\usepackage{amscd}
\usepackage{epsfig}
\usepackage{amssymb}
\usepackage{tabularx}

\def\mc{{\cal M}_c}

\def\tr{{\rm Tr\, }}
\def\c{{\Bbb C}}
\def\c{{\mathbb{C}}}

\def\dd{\hbox{\kern0.3em/\kern-0.7em /\kern0.5em}}
\def\lie#1{{\rm Lie}\left( #1 \right)}

\def\a{\alpha}

\def\g{\gamma}
\def\e{\epsilon}

\def\d{\delta}
\def\t{\theta}
\def\hp{\hat{\phi}}

\def\t#1{{\Bbb T}_{#1}}
\def\1#1{{\Bbb N}_{#1}}

\def\tr#1{tr_{#1}\;}

\def\g{\mathfrak g}
\def\h{\mathfrak h}
\def\r{\mathfrak r}

\def\rg{r_g {\mathbb I}}
\def\Z{\Bbb Z}

\def\ssqr#1#2{{\vbox{\hrule height #2pt
\hbox{\vrule width #2pt height#1pt \kern#1pt\vrule width #2pt}
\hrule height #2pt}\kern- #2pt}}

\newcommand{\drawsquare}[2]{\hbox{%
\rule{#2pt}{#1pt}\hskip-#2pt
\rule{#1pt}{#2pt}\hskip-#1pt
\rule[#1pt]{#1pt}{#2pt}}\rule[#1pt]{#2pt}{#2pt}\hskip-#2pt
\rule{#2pt}{#1pt}}

\newcommand{\Yfund}{\raisebox{-.5pt}{\drawsquare{6.5}{0.4}}}
\newcommand{\Ysymm}{\raisebox{-.5pt}{\drawsquare{6.5}{0.4}}\hskip-0.4pt%
        \raisebox{-.5pt}{\drawsquare{6.5}{0.4}}}
\newcommand{\Yasymm}{\raisebox{-3.5pt}{\drawsquare{6.5}{0.4}}\hskip-6.9pt%
        \raisebox{3pt}{\drawsquare{6.5}{0.4}}}

\newcommand{\Yfoura}{\raisebox{-3.5pt}{\drawsquare{6.5}{0.4}}\hskip-6.9pt%
        \raisebox{3pt}{\drawsquare{6.5}{0.4}}\hskip-6.9pt
        \raisebox{9.5pt}{\drawsquare{6.5}{0.4}}\hskip-6.9pt
        \raisebox{16pt}{\drawsquare{6.5}{0.4}}}
\begin{document}
\tighten

\preprint{\vbox{
\hbox{hep-th/0207151}
}}
\title{Discrete anomalies and the null cone 
of SYM 
theories}
\author{Gustavo Dotti}
\affiliation{FaMAF, Universidad Nacional de C\'ordoba,\\
Ciudad Universitaria, 5000, C\'ordoba, ARGENTINA}
\email{gdotti@fis.uncor.edu}
\date{April 2003}

\begin{abstract}
A stronger version of an anomaly matching theorem (AMT) is proven
that allows to anticipate the matching of continuous as well as  discrete 
global anomalies. The AMT
 shows a connection between anomaly matching and the geometry 
of the null cone of SYM theories. Discrete symmetries are shown 
to be broken at the origin of the moduli space in Seiberg-Witten theories.

\end{abstract}


\maketitle


Global symmetries play an important role in the study of 
supersymmetric gauge theories. 
In particular, 't Hooft condition \cite{th} that the massless fermions 
in the low energy theory have the same 
continuous  global symmetry anomalies as the 
fundamental fields
is  so restrictive,  
that  gives us a strong confidence on a proposed low energy spectrum 
if  this test is passed.  Analogous  conditions 
for  discrete  symmetries were given in \cite{dam} and references 
therein. Anomaly matching can be used to set necessary conditions 
to decide if the classical moduli space $\mc$ of a SYM theory correctly 
describes the set of vacua and their  low energy
 massless spectrum in the quantum regime.
Computing anomalies at a point of $\mc$, however, 
is a difficult task that implies finding the basic invariants and 
their constraints, linearizing constraints at the desired point, 
and decomposing  the resulting tangent space into 
irreducible representations  of the flavor group.
 Theorems I and II in \cite{dm1}
allow to anticipate the outcome of this test 
after a  simple inspection of a sample of  points in the elementary field
 space that are D-flat and completely break the gauge group \footnote{A complexified 
gauge orbit is closed, as required in Theorem I in \cite{dm1}, if and 
only if is the orbit of a D-flat point.}.
In this letter we prove a stronger Anomaly Matching Theorem (AMT)  
improved to: (i)  anticipate the matching of {\em discrete} anomalies 
as well as continuous ones,  
(ii) allow non D-flat points in the sample set of elementary fields, 
(iii) allow field configuration that do not break the gauge group
completely. Condition (ii) is very useful when dealing with theories 
with unconstrained basic invariants, because the matching of 
the anomalies of the full global symmetry group 
can be checked by looking at a single point, a suitable 
non zero  elementary field configuration above the origin 
of moduli space. A point like that   is necessarily non D-flat.
Condition (iii) is useful  
to understand why anomalies match in theories such as $SO(N)$ with $N-4$ 
vectors.\\
Throughout the paper we use the following terminology and notation:
$\phi \in \c^n = \{ \phi \}$ denotes a spacetime constant  
configuration of the elementary matter chiral fields.
 $G$ is the gauge group, $\rho$ its representation 
 on $\{ \phi \}$,  $\rho = \oplus_{i=1}^k F_i \rho_i$ its decomposition 
into irreducible representations.
 The classical flavor group is $\hat F = SU(F_1) \times \cdots \times 
SU(F_k) \times U(1)_R \times U(1)^k$, anomalies break 
 the $U(1)^k$ piece 
 down to $U(1)^{k-1} \times \Z_{\mu}$, $\mu = \sum_i F_i \mu_i$ 
the ( properly normalized \cite{dam} ) Dynkin index of $\rho$. The resulting 
quantum flavor group is called $F$, all
the elementary fields have charge one under  $\Z_{\mu}$. 
$\hp^{\a}(\phi), \a=1,...,s$ is a basic set of homogeneous, holomorphic  
$G$ invariant polynomials  on $\c^n$, which,  
being holomorphic,  are also invariant 
under the action of the complexified gauge group $G^c$.
The level sets $\hp(\phi) = \hp_o$ of the map $\hp:~\c^n \to \c^s$ 
 are called {\em fibers}, the fiber through 
$\phi$ being ${\cal F}_{\phi}  = 
\hp^{-1}(\hp(\phi))$. Due to the $G^c$ invariance of the map 
$\hp: \c^n \to \c^s$, 
fibers contain complete $G^c$ orbits. 
There is precisely one $G$ orbit of $D-$flat points in every fiber 
\cite{plb}, then,  
for theories with zero 
superpotential, 
the classical moduli space $\mc \equiv  \{D-\text{flat points}\}/G 
= \hp( \c^n ) \subseteq \c^s$.
A particularly important fiber we will be dealing with is the one through 
$\phi=0$, ${\cal F}_0 = \{ \phi \in \c^n | \hp(\phi)=0 \}$, called 
the {\em null cone}. The  $G$ orbit of $D-$flat points in
the null cone is $\{ 0\}$.
If the basic invariants $\hp^{\a}(\phi), 
\a=1,...,s$ are 
algebraically independent  then $\mc = \hp(\c^n)=\c^s$. If 
 they are constrained by polynomial relations 
$p_r(\hp(\phi)) \equiv 0$, then $\mc=\hp(\c^n)=\{\hp \in \c^s|p_r(\hp)=0\}$. 
The tangent space of $\mc$ at $\hp$, denoted $T_{\hp}\mc$, 
is the linear set of allowed $\d \hp's$  
 obtained by linearizing at $\hp$ the 
constraints  $p_r(\hp)=0$.
The differential at $\phi$ of $\hp:\c^n \to \mc$,  
denoted $\hp'_{\phi}$,  
 is a linear map from the tangent space $T_{\phi} \c^n \simeq \c^n$ 
into 
the tangent space at $\hp(\phi)$ of the moduli space: 
$\hp'_{\phi}: \c^n \to T_{\hp(\phi)} \mc$, 
$\hp'_{\phi}: \delta \phi^i \to \delta \hp^{\a} = 
(\partial \hp^{\a}/\partial 
\phi^j)  \delta \phi^j$.  
't Hooft's condition  states  that the global and gravitational anomalies
of the unbroken symmetries at a given vacuum, computed 
in the space of  massless elementary fermions 
(or UV, for ultraviolet, 
as in \cite{dm1,dm2}) should match the corresponding anomalies 
in the low energy theory 
(or IR, for infrared sector),
 i.e., given any three global symmetry unbroken 
generators 
\begin{eqnarray} \label{amc}
\tr{IR} \h_A \{\h_B,\h_C\} = \tr{UV} \h_A \{\h_B,\h_C\}  \\
 \label{gamc} \tr{IR} \h_i = \tr{UV} \h_i, i=A,B,C \;\;
\end{eqnarray}
 Discrete symmetries classify into two types, 
and only type~I anomalies are required to match \cite{dam}. Type~I anomalies 
are $SU^2 \Z_{\mu}$, which should only match mod $\mu$, and the 
gravitational $\Z_{\mu}$, which has to match mod $\mu/2$ \cite{dam}.
The UV space on the rhs of (\ref{amc}) and (\ref{gamc}) is the complex vector 
space 
$\{\phi\}$ of elementary chiral matter fields, plus the gaugino space 
$\lie {G^c}$ (which only contributes in the case of $R$ symmetries). 
We are interested in checking if $\mc$ gives a  correct IR description, 
i.e., if  IR $= T_{\hp}\mc$ (plus leftover gauginos if the theory is 
in a Coulomb phase) passes 't Hooft test. This has implications for 
theories with  quantum modified moduli spaces \cite{dm1,dm2}.
 Theorems I and II 
in \cite{dm1}
allow us to anticipate (without even finding the invariants to construct
 $\mc$)
the matching of continuous global anomalies at $\hp_o$  if 
there is a D-flat point $\phi$ that completely breaks $G$ 
and satisfies $\hp(\phi)=\hp_o$. Since there is no such a D-flat 
point over the origin $\hp_o=0$ of $\mc$, to explain
anomaly matching at the origin of, e.g,  theories with an 
affine moduli space (AMS) \cite{prl} or s-confining theories \cite{sc},
 a set $\phi_j$ of D-flat points that completely 
break $G$ is used,  such that anomaly matching for the unbroken  symmetry 
groups $F_i$ imply matching for $F$ at the origin \cite{dm1,dm2}. 
The results in \cite{dm1,dm2} 
were  used to prove anomaly matching at every point of $\mc$ for 
 s-confining theories and theories with a quantum modified moduli space, 
and also to show 
that anomaly matching in dual theories is a consequence  of the 
similarities in their chiral rings \cite{dual}.
They do not allow, however, to test discrete anomaly matching and 
see if it is a truly independent test. The anomaly matching theorem 
(AMT) below overcomes this difficulty, and can also be applied at 
points $\phi$ that maximally break $G$, even if $G$ is not 
completely broken, and most important, even if $\phi$ is not D-flat.
As an example, $\phi$ could be a point in the null cone (i.e., 
$\hp(\phi)=0$), where $F$ is unbroken, and be used 
to anticipate  full anomaly matching in AMS theories, or Seiberg-Witten (SW) 
 theories \cite{sw}.
As we will see, computations do not simplify when testing discrete anomalies, or 
when the gauge group is not completely broken.
However,  interesting relations  arise between  the geometry of the null-cone 
and  anomaly matching at the origin of $\mc$. \\

\noindent
{\bf Anomaly Matching Theorem:} (compare to Theorems I and II in 
 ref \cite{dm1}) Assume that  $G$ is semisimple, 
$G^c \phi_o$ has maximal dimension and $\hp(\phi_o)$ is a smooth point 
of $\mc$. Let  $\h_A,\h_B$ and $\h_C$ be any three generators  of 
the unbroken global symmetry 
subgroup at $\hp(\phi_o)$, 
 with the first 
 $k$  of them ($k=0,1,2,3$)
equal 
to $\h={\Bbb I}$, the generator of the anomalous 
$U(1) \supset \Z_{\mu}$. \\

There are 
generators  $\g_A,\g_B$ and $\g_C$ of $G^c$ such that 
$(\h_i+\g_i)\phi_o=0$ for $i=A,B,C$. Furthermore, 
the UV - $T_{\hp(\phi_o)}\mc$ 
gravitational anomaly mismatch is:
\begin{equation} \nonumber
\tr{UV} \h_i  = \tr{(T_{\hp(\phi_o)}\mc)} \h_i 
- \tr{\lie{{G^c}_{\phi_o}}}Ad_{\g_i} 
\end{equation}
and the flavor anomaly mismatch is 
\begin{multline} \label{amt}
\tr{UV} \h_A \{\h_B,\h_C\} =
\tr{(T_{\hp(\phi_o)}\mc)} \h_A \{ \h_B,\h_C\}  \\
- \tr{\lie{{G^c}_{\phi_o}}}  Ad_{\g_A} 
\{Ad_{\g_B},Ad_{\g_C}\} \\
- k \; \tr{\{\phi\}} \{\g_B,\g_C\}
\end{multline}
where $Ad_{\g_i} $ must be replaced 
with $ {Ad_{\g_i}}-r_g {\Bbb I}$ 
if $\h_i$ is an $R$ symmetry ($r_g$ is the gaugino 
$U(1)_R$ charge \footnote{Usually set equal to one, it may be assigned a
different value to avoid non integer $U(1)_R$ charges. This is relevant when 
computing mixed discrete - $U(1)_R$ anomalies.}.)\\

Before proving the theorem we will show some applications.\\
{\sc Theories with D-flat points that break $G$ completely:}
This is the case studied in \cite{dm1,dm2}.
If  $\phi_o$  is D-flat and 
breaks $G$ completely  then 
 also breaks $G^c$ completely \cite{gatto}. Furthermore, 
  $\hp(\phi_o)$ is in the 
principal stratum of $\mc$, and so  is 
smooth \cite{plb}. 
The  AMT eq.(\ref{amt}) can  then be applied at $\phi_o$. 
Since $\lie { {G^c}_{\phi_o}}$ is trivial, according to the AMT
 continuous anomalies 
 will match 
 between the UV and $T_{\hp(\phi_o)}\mc$. 
This argument holds for  {\em every} point $\hp$ in the principal  
stratum of theories where the gauge group can be completely broken, 
among 
which are all SYM theories with matter in gauge representations 
with Dynkin index greater than the index of the adjoint  \cite{ela}. 
These facts  can be used to simplify the proofs   in \cite{dm1,dm2} 
that anomalies match at every point of the moduli space  
for s-confining theories and 
theories with 
a quantum modified moduli space; and  also the proof in 
\cite{dual} that  matching  in dual theories 
is a consequence of the similarities of the chiral rings of the duals, rather than 
an independent duality test. \\

{\sc Theories with unconstrained basic invariants:}
We show some applications of the AMT that require the stronger 
version given above. 
Instead of looking at $\hp$ in the principal 
stratum, we explore the opposite situation: $\hp=0$. 
We  would like to 
understand 
 why anomalies match at the origin $\hp=0$ 
of the moduli space of theories such as 
the AMS theories  in \cite{prl},
or the   Seiberg-Witten theories \cite{sw}. 
These are examples of theories with  unconstrained
basic invariants,  for which   $\mc$ is a vector space, and  $\hp=0$ 
a smooth point. The  AMT   can be applied at $\hp=0$ 
if there is a point 
$\phi_o$ 
in the null-cone that maximally breaks $G^c$, i.e.
\begin{equation} \label{ncp}
\hp(\phi_o) = 0,  \;\; \text{ dim }\; G^c \phi_o = d \; \text{ (maximal)}
\end{equation}
If such a point exists, the anomaly mismatch
 between the basic invariants and the UV is given by 
 eq.(\ref{amt}). In particular, if  $d = d_G \equiv \text{dim } G$ 
all  continuous  anomalies 
must match. \\
Eight  out of the eleven AMS theories in table~I of  \cite{prl}
have matter content in irreducible representations of the gauge group.
 Irreducible representations of simple 
gauge groups with unconstrained basic invariants 
share  the rare  property that all of their fibers 
 have the same dimension and contain a 
finite number of $G^c $ orbits \cite{bible}. Thus,  
  the dimension of a
 fiber $f$ equals the maximum dimension of a $G^c$ orbit in it, 
say $d_f$, and, 
since all fibers have the same dimension, it must be $d_f=d$ for all $f$, in particular, 
for the null cone. We conclude that  
eq.(\ref{ncp})  has a solution. 
Besides being irreducible, the gauge representations in 
Theories T8 through T11 of Table~I in \cite{prl} 
have Dynkin index $\mu$ greater than the adjoint index $\mu_{adj}$, then 
dim $G^c \phi_o=d_G$ for $\phi_o$ satisfying (\ref{ncp}). Applying the AMT 
at $\phi_o$ continuous anomaly matching at the origin follows. 
The  existence of a maximal dimension $G^c$ 
orbit in the null cone implies continuous anomaly matching!
Why do continuous anomalies match for the other AMS theories? 
According to the AMT, because 
the restricted$Ad_{\g}$ representation of (\ref{amt}) 
is anomaly free. As an example consider the theory with 
 $G=SO(2n+k)$ and  $n$ 
vectors, collected in a $(2n+k) \times n$ matrix $\phi$.
The flavor group 
is $SU(n) \times U(1)_R \times Z_{2n}$, 
the scalar piece of $\phi$ transforms 
as $(\Yfund,(2-n-k)/n,1).$ 
The point 
\begin{equation} \label{sopoint}
\phi_0 = 
\left( \begin{array}{c} {\Bbb I}_{n \times n} \\ i{\Bbb I}_{n \times n}\\
0_{k \times n} \end{array} \right).
\end{equation}
satisfies eq.(\ref{ncp}). The Lie algebra of the unbroken gauge   group at 
$\phi_o$ 
is spanned by matrices of the form
\begin{equation} \label{iso}
\left( \begin{array}{ccc} A & iA & B \\ iA & -A & iB \\-B^T & -iB^T & A' 
\end{array} \right)
\end{equation}
with $A$ an antisymmetric $n \times n$ block and $A'$
 an antisymmetric $k \times k$ block. Given 
$\h$ in  $\lie {SU(n)}$, 
$\h=s+ia$
($s$ real symmetric $a$ real antisymmetric),  
a $\g_{\h}$ satisfying $(\g_h+\h)\phi_o=0$,
 (predicted by  the AMT)  can be chosen as 
\begin{equation} \label{su}
\g_{\h} =  -\left( \begin{array}{ccc} ia & -is & 0 \\ is & ia & 0
\\0&0&0 \end{array} 
\right) 
\end{equation}
 Similarly,  $\g_{\r}= 
-r \g$ and $\g_Z = -\g$, where $r=(2-n-k)/n$ is the $r-$charge of the 
scalar fields and 
\begin{equation} \label{gz}
\g = \left( \begin{array}{ccc} 0_{n \times n}& -i{\Bbb I}_{n \times n} & 
0_{n \times k}\\
 i{\Bbb I}_{n \times n} &  0_{n \times n} &  0_{n \times k} \\
0_{k \times n} & 0_{k \times n} & 0_{k \times k} \end{array} \right), 
\end{equation} 
Under the  $SU(n) \times U(1)_R \times \Z_{2n}$  
$Ad$ representation ($Ad-1$ for the $U(1)_R$ generator) 
that enters eq (\ref{amt}), $A$ is 
a $ (\Yasymm,-1+2(n+k-2)/n,-2)$, $B$ 
 a $(k \Yfund,  -1+(n+k-2)/n, -1)$ and 
$A'$ a $(1,-1,0)$.  
This representation  happens to be anomaly free 
precisely when  $k=4-n$, i.e., when  the number of  
vectors is four less than the number of colors, as expected. 
This is why global anomalies match in this case. 
Back to  theories with $d=d_G$,  we may ask what the AMT 
tells us about {\em discrete} symmetries.
To test the matching of discrete anomalies we 
need to explicitely {\em find} $\phi_o$ and $\g_{Z}$,  as done 
done above (eqns (\ref{sopoint})  and (\ref{gz})), it is no longer 
sufficient to know that such a $\phi_o$ exists. Although this makes 
the AMT computationally useless (it is certainly easier to go through 
the usual steps to verify anomaly matching), an interesting geometrical 
picture arises.
Solving eq.(\ref{ncp}) for theories with invariants 
of high degree is a very difficult problem, we would like to show 
a way around it. A recipe to find points in the null cone is 
given in \cite{bible}: fix a Cartan algebra, obtain 
the weight decomposition 
$\phi = \sum_{\lambda} \phi_{\lambda},\phi_{\lambda}\neq 0$, 
and construct  the convex hull spanned by the weights $\lambda$. If the hull 
 does not contain the origin then $\phi$ is in the null cone. 
Now  add the requirement 
that  $\phi$ break $G^c$ completely. This condition is guaranteed if
we  make sure that under no root translation will  
 all  the $\lambda$'s in the weight decomposition of $\phi$
  ``drop off'' the weight diagram, and  that 
two different root translations will not leave exactly the same weight 
spaces occupied. For example, in the AMS 
theory $G=SO(5)$ with matter in the $\Ysymm$ we readily see from figure 1 below 
that points  containing the weights  $\e_1, \e_1 \pm \e_2, 2\e_1, 
\e_2$ and $2 \e_2$ are in the null cone 
(any Weyl rotation of this set would equally work). 
 We have made the standard 
choice of Cartan generators ($\frak C_i$ has a
 Pauli $\sigma_2$ matrix in the $i-th$ $2 \times 2$ diagonal block)
and weights,  $e_j(\frak C_i) = \d_{ij}$.
Since  $\e_1-\e_2$ and $\e_2$ are not simultaneously 
drop off the diagram  under  root translations, vectors in $V_{\e_1-\e_2} 
\oplus V_{\e_2}$ completely break $SO(5,\c)$, and so are 
particular solutions of eq.(\ref{ncp}), as the reader may check.
 Is there any other  property to require 
from $\phi_0$? We may ask that $\g_Z (\propto \g_{\r}) $ in the AMT 
belongs to the Cartan 
subalgebra. If this is the case,  then 
$\g_Z \phi_o = - \phi_o$
 reads  $\lambda(\g_Z)=-1$ 
for all $\lambda$ in the weight decomposition of $\phi_0$. 
 This 
defines a hyperplane in weight space,  the weights in 
$\phi_0$ must lie in this  hyperplane if we want $\g_Z$ 
in the Cartan subalgebra. In the $SO(5)$ theory with a $\Ysymm$ 
 a possible choice is  $\phi_0 \in  V_{ \e_2} \oplus
V_{\e_1-\e_2}$,  with 
$\g_Z=-(2 \frak C _1 + \frak C _2) $.
\begin{figure}[h]
\epsfxsize=8cm
\centerline{\epsffile{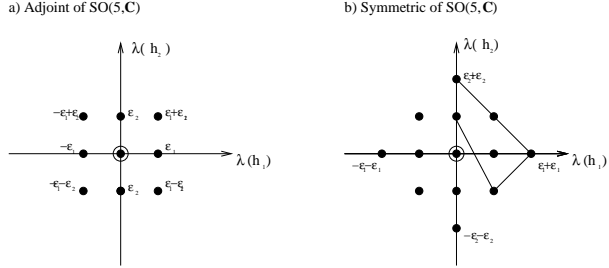}}
\caption[]{Weight diagram of the adjoint (a) and symmetric  (b) of $SO(5,\c)$ 
showing  the convex hull spanned by $\e_1, \e_1 \pm \e_2, 2\e_1, 
\e_2$ and $2 \e_2$} 
\label{so5wd}
\end{figure}
Once we have the $\g_i$'s of the AMT, we may apply eq.(\ref{amt}).
Note that  the mismatch of anomalies involving $U(1)_R$ and $\Z_{\mu}$ 
will be  proportional to $\tr{\{ \phi \}}{{\g_Z}^2} = (\mu/\mu_{adj})
\tr{adj}{{\g_Z}^2} \propto 1/D^2$, 
$D$ the distance to the origin of weight space of the hyperplane containing 
the weights of $\phi_o$. We conclude that, for the
 theories under consideration: 
(i)  the matching of continuous anomalies 
 is a consequence of the existence of an orbit  $G^c\phi_o$ 
  of maximal 
dimension in the null cone,   the weights of $\phi_o$ 
can be chosen lying on a  hyperplane in weight space, and (ii)  the distance of 
this hyperplane  to 
the origin gives the discrete anomaly mismatch. 
This is the anomaly matching - geometry interplay 
refered to above.\\
Theories 
in a Coulomb phase also  have a classical
 moduli space spanned by unconstrained  basic invariants.
 We will concentrate on the SW theories with one flavor 
of matter in the adjoint (of a simple gauge group $G$). The flavor 
flavor group is $U(1)_R \times \Z_{\mu_{adj}}$, 
the unbroken gauge subgroup  at a $D-$flat point
that maximally breaks $G$ 
 is $U(1)^r$, $r$ the rank of $G$. 
Continuous symmetry anomalies are known to match between the 
elementary fields and the  unconstrained moduli  if the 
unbroken gauginos in $\lie {U(1)^r}$, which transform non trivially under 
$U(1)_R$, are added to the moduli.
The matching of $U(1)_R$ symmetries follows readily from   the AMT,
eq.(\ref{amt}), when applied at a  $D-$flat point $\phi_o$ that breaks $G$ to 
$U(1)^r$. Since $U(1)_R$ acts trivially on the scalar matter fields, 
we can choose 
$\g_{\h}=0$ ($\h$ the generator of $U(1)_R$),  then  $Ad_{\g_{\h}}=0$ 
and  (\ref{amt})  precisely says that 
$U(1)_R$ anomalies match if the light, unbroken 
$\lie {U(1)}^r$ gauginos are assigned 
charge one and added to the moduli.  Regarding discrete symmetries, 
they can be treated exactly as done for the $SO(5)$ theory above. 
A point satisfying (\ref{ncp}) can be explicitely found 
using elementary Lie algebra facts. 
It is found that type~I \cite{dam} discrete anomalies only match 
for the even rank, simply laced $G: A_{2n}, D_{2n}, E_6$ 
and $E_8$. This implies that $Z_{\mu}$ must be broken at the origin 
of the other SW theories \cite{dam}, a fact that is not obvious,   
 since the vev's of all basic 
invariants are zero. A possible explanation is that the microscopic 
fields that span the Cartan subalgebra of the matter fields in the adjoint 
do get a nonzero vev, showing that $Z_{\mu}$ is actually broken \footnote{
I thank Witold Skiba for suggesting this possibility.}.
In the $SU(N)$ case, the Cartan subalgebra is the subspace 
of the diagonal Lie algebra matrices 
defined by $\sum_{k=1}^N a_k=0$, and the vev's $<a_k>$ 
are given by \cite{sw,kly}
\begin{equation} \label{vev}
<a_i> = \oint_{{\cal \gamma}_i} \lambda_{SW} dx,
\end{equation}
where $\gamma_i$ are the $N$ branch cuts of the complex 
function $y=f(x)$ defined by \cite{kly}
\begin{equation} \label{curve}
y^2 = k(x) \equiv \left( \sum_{\a=0}^N s_{\a}x^{N-\a} \right)^2
 - \Lambda^{2N},
\end{equation}
and the Seiberg-Witten one form is 
\begin{equation} \label{swf}
 \lambda_{SW} \propto \left( \sum_{\a=0}^N (N-\a)s_{\a}x^{N-\a} 
\right) \frac{dx}{y}
\end{equation}
Here $s_0=1,s_1=0$ and the remaining ${s_{\a}}'s$ are the vev's of 
basic invariants, related to the standard invariants 
$\hp^{\a}=\text{tr } {\phi^{\a}}$ through 
$$r s_r+ \sum_{\a=0}
^r s_{r-\a} \hp^{\a}=0, \hspace{1cm} r=1,2,3,...$$
The $2N$ zeroes of $k(x)$ in (\ref{curve}) are of the form 
$x=g_i(s,\pm \Lambda^n), i=1,...,N$, where the $g_i$ can be
unambiguously defined if $|s| \gg |\lambda|$, then the cuts ${\cal C}_i, 
i=1,...,N$  of 
$$y(x)= \prod_i [\sqrt{x-g_i(s,\lambda^n)} \sqrt{x-g_i(s,-\lambda^n)}]$$ 
can be chosen with ${\cal C}_i$ 
the segment 
from $g_i(s,\lambda^n)$ to $g_i(s,-\lambda^n)$.
For the vacuum at the origin of the moduli space, the roots of 
$k(x)$ are $\omega^{i-1}\Lambda, i=1,...,N$, 
$\omega=e^{i\pi/N}$, and  it is  not 
obvious 
how  the $N$ roots of $1$ pair to the $N$ roots 
of $-1$. From the  the observation in \cite{kly} that 
a rotation $\Lambda^{2N} \to e^{2\pi i t}\Lambda^{2N},t\in [0,1]$ 
transforms a root of $k$ into its pair, we conclude that the 
cut ${\cal C}_i$  links $\omega^{2i-2} \Lambda$ to $\omega^{2i-1}\Lambda$, 
$i=1,...,N$, then 
\begin{equation} \label{vev2}
<a_i> \;  \propto \; \int_{\omega^{2i-2} \Lambda}^{\omega^{2i-1}\Lambda} 
\frac{x^N}{\sqrt{x^{2N}-\Lambda^{2N}}} dx
\end{equation}
A change of the integration variable to $z=\omega^2 x$ shows that 
$<a_{i+1}> = \omega^2 <a_i>$, then $<\sum_{i=1}^n a_i> =0$, as expected.
The change of integration variable $z=x^n$ in (\ref{vev2})
\begin{equation} \label{vev3}
<a_1> \propto \int_{-\Lambda^{N}}^{\Lambda^{N}} 
\frac{z^{1/N}}{\sqrt{z^2-\Lambda^{2N}}}
\end{equation}
 may suggest the $<a_i>=0$ if $N$ is odd, explaining the discrete anomaly
 matching at the origin of the odd $N$  theories.
However this  is not correct, 
the branch of $z^{1/N}$ in (\ref{vev3}) is not the
one taking  real values for negative $z$,  
the integrand is not odd, and the $<a_i>$ do not vanish. 
We have constructed the correct integrand (reproducing the 
desired branch cuts) and evaluated  numerically the integrals 
defining $<a_1>$ for $SU(N)$  for the first few $N'$s. We have found 
that 
$<a_1> $ does {\em not} vanish, even for odd N. 
Repeating this calculation for other simple groups, such 
as the exceptional groups, 
is much more difficult, due to their complex branch structure. 
Our results,  however,  seem to indicate that 
anomaly matching at the origin of the even rank, simply laced SW theories, 
is accidental.\\

\noindent
{\bf Proof of the AMT:} Since $G$ is semisimple, $G^c \phi_o$ has 
maximal dimension and $\hp(\phi_o)$ is smooth, the differential 
$\hp'_{\phi_o}: \{ \phi \} \to T_{\hp(\phi_o)}\mc$ is  onto \cite{knop},
 and (using dim $\mc = $ dim $
 \{ \phi \} - $ dim $G$) has kernel $\lie {G^c} \phi_o \equiv \t{\phi_o}$. 
If  $\h \in \lie {{\hat F}_{\hp(\phi_o)}}$, then $0 = \h \hp(\phi_o) = 
\hp'_{\phi_o} \h \phi_o$. Since   $\h \phi_o \in $ ker 
$\hp'_{\phi_o} =  \t{\phi_o}$, there is a $\g_{\h} \in \lie {G^c}$ 
such that $\g_{\h}+\h \in \lie{(G^c \times {\hat F})_{\phi_o}}$.
 $\g_{\h}$  is not unique if $\lie {{G^c}_{\phi_o}}$ is non trivial, 
and so there is no 
``star'' flavor 
representation under which  $\{ \phi \}$ breaks into $\t{\phi_o}$ 
plus an invariant complement,  as in the proof of Theorem II in \cite{dm1}. 
It can easily be checked that 
 $\t{\phi_o}$ is invariant under 
 $(G^c \times {\hat F})_{\phi_o}$, 
 but this group may be non reductive if  $\phi_o$
is not D-flat, and this implies  that 
$\t{\phi_o}$ may not have an invariant complement (as an example, 
consider $\phi_o$ of eq.(\ref{sopoint})).  
The way out of this problem is to work with 
the {\em quotient} vector space $\{ \phi \} / \t{\phi_o}$, where a
$(G^c \times {\hat F})_{\phi_o}$ action is well defined, and the 
following diagram commutes \footnote{Given an onto linear map $O:V \to 
W$ with kernel $V_o$, we call $[O]:V/V_o \to W$ the induced isomorphism. 
If $O:V\to V$ is linear and $V_o  \subset V$ an invariant subspace, $[O]$ 
denotes the 
induced linear map $V/V_o \to V/V_o$}:
\begin{equation} \label{cd}
\begin{CD}
\{ \phi \}/ \t{\phi_o}  @>{[\hp'_{\phi}]}>> T_{\hp(\phi_o)}\mc \\
@V{[gh]}VV                 @VV{h}V \\
\{ \phi \}/ \t{\phi_o}   @>[\hp'_{\phi_o}]>> T_{\hp(\phi_o)}\mc
\end{CD}
\end{equation}
Now 
consider the map $t: \lie{G^c} \to \t{\phi_o}$ given by $t({\mathfrak {g}}) = 
{\mathfrak {g}} \phi_o$. This map is onto and has kernel 
$\lie {{G^c}_{\phi_o}}$. 
If $(g,h) \in (G^c \times {\hat F})_{\phi_o}$ then 
$gh {\mathfrak {g}} \phi_o = 
gh {\mathfrak {g}} (gh)^{-1} \phi = (g{\mathfrak {g}}g^{-1}) 
\phi_o$, and also   $g  \lie {{G^c}_{\phi_o}} g^{-1} \subseteq 
\lie {{G^c}_{\phi_o}}$ . We have a situation analogous 
 to that leading to the diagram~(\ref{cd}):
\begin{equation} \label{cd2}
\begin{CD}
\lie {G^c} /\lie {{G^c}_{\phi_o}}   @>[t]>> \t {\phi_o} \\
@V[Ad_g]VV                 @VV(g , h)V \\
\lie {G^c} /\lie {{G^c}_{\phi_o}}   @>[t]>> \t{\phi_o} 
\end{CD}, 
\end{equation}  
Now let $\h_i, i=A,B,C$ be three  non $U(1)_R$ generators 
of ${{\hat F}}_{\hp(\phi_o)}$. 
The differential version of (\ref{cd}) reads 
\begin{equation} \label{difcd1}
{\h_i}_{\mid_{T_{\hp(\phi_o)\mc}}}= 
[ \hp'_{\phi_o}] \left[{[\g_i+\h_i]}_{\mid_{\{ \phi \} /  \t{\phi_o}}} \right]
 {[\hp'_{\phi_o}]}{}^{-1}
\end{equation}
and  that of (\ref{cd2}) is 
\begin{equation} \label{difcd2}
{(\g_i+\h_i)}_{\mid_{\t{\phi_o}}}=
[t] \left[{Ad_{\g_i}}_{\mid_{\lie {G^c} /\lie {{G^c}_{\phi_o}}}}\right]
[t]^{-1},
\end{equation}
If $O:V \to V$ is a linear operator that leaves the subspace
 $W \subset V$ invariant, 
 $\tr{V/W}[O] = \tr{V}O-\tr{W}O$. This, together with  (\ref{difcd1}), 
(\ref{difcd2}), the facts that 
   any $G$ representation is traceless for  $G$ semisimple, and that 
the $G$ action on $\{ \phi \}$ is free of anomalies,  
imply
\begin{equation} \nonumber
\tr{T_{\hp(\phi_o)}\mc} \h_i =  \tr{\{ \phi \}} \h_i  + 
\tr{\lie{{G^c}_{\phi_o}}}Ad_{\g_i}. \label{grav}
\end{equation}
and 
\begin{multline}\label{amfpua} 
\tr{T_{\hp(\phi_o)}\mc} \h_A \{ \h_B,\h_C\}-\tr{\{ \phi \}}
 \h_A \{\h_B,\h_C\} = \\
 \tr{\lie{{G^c}_{\phi_o}}} Ad_{\g_A} \{Ad_{\g_B},Ad_{\g_C}\} 
+ \tr{\{ \phi \}}  \g_A \{\g_B,\h_C\} \\
+ \tr{\{ \phi \}} \g_A \{\h_B,\g_C\}+\tr{\{ \phi \}} \h_A \{\g_B,\g_C\} 
\end{multline}
The last three terms vanish for non anomalous ${\hat F}$ generators, 
and give the $k$ term of (\ref{amt})  when $k$ of the $\h_i$'s equal 
 the anomalous $\h= {\Bbb I}$ that generates $U(1)_A \supset \Z_{\mu}$. 
Anomalies involving 
$\h \in \lie{U(1)_R}$ must be computed    
using the fermionic $\tilde \h  \equiv  \h -r_g{\Bbb I}$ charge matrix.
 We leave it for the reader to check that calculations go through if 
we replace $Ad_{\g_i} $
with ${Ad_{\g_i}}-r_g {\Bbb I}$, use the facts that $\tilde \h$ 
(instead of $\h$) is anomaly free, and that $UV$ gets enlarged 
to $\{ \phi \} \oplus \lie {{G^c}_{\phi_o}}$ for cubic $U(1)_R$ anomalies. 
 The AMT then follows.\\
This work was supported by Conicet and Secyt-UNC. I thank Witold 
Skiba for clarifications about theories in a Coulomb phase.

\end{document}